# Experimental Study of the Effect of External Signal on Microwave Oscillations in a Nonrelativistic Electron Beam with Virtual Cathode

**Yu. A. Kalinin and A. E. Hramov***

*Saratov State University, Saratov, Russia*
*\*e-mail: noios@sgu.ru; aeh@nonlin.sgu.ru*

**Abstract**—The effect of an external harmonic signal on the characteristics of microwave generation in a nonrelativistic electron beam with virtual cathode (VC) formed in a static retarding electric field (low-voltage vircator system) has been experimentally studied. A significant increase in the vircator generation power is observed when the frequency of the external signal is close to the frequency of VC oscillations. At large detunings, a broadband chaotic generation is observed.

Microwave electronic devices with the active medium in the form of an electron beam with a virtual cathode (VC) are promising sources of microwave radiation on various power output levels [1–7]. Recently, we proposed and studied [6–8] a new scheme of oscillator with a VC (vircator), which employs an intense nonrelativistic electron beam (i.e., a beam with a microperveance $p_\mu > 3$ μA/V$^{3/2}$ [9]). In order to form a VC in the electron beam, this system employs a scheme with an additional retardation of electrons. According to this scheme (called a low-voltage vircator), a nonstationary oscillating VC is formed at the expense of strong retardation of electrons in the drift region [6], which makes possible the generation of both single-frequency and broadband microwave signals using electron beams with small total currents and low densities [7]. In such regimes, it is possible to study in much detail the physical processes in electron beams with VCs. The low-voltage vircator system is also of interest as a controlled source of medium-power broadband chaotic signals in the centimeter and millimeter wavelength ranges [8].

An important problem in the study of vircators is related to an analysis of the influence of external signals on the generation characteristics (spectrum, generation bandwidth, power) in the electron beam with a VC. This problem is of special interest in view of the possible use of vircators as microwave amplifiers. Experimental and theoretical investigations of the effect of external signals on the vircator operation have been reported in [10–13] (see also review [5]). Recently, we reported [14] on the phenomenon of synchronization of VC oscillations in a nonrelativistic electron beam under the conditions of additional retardation of electrons. However, the general pattern of nonautonomous dynamics in electron-wave systems with VCs is still not completely clear.

This Letter presents the results of an experimental investigation of the effect of an external harmonic microwave signal on the generation characteristics of a low-voltage vircator.

The low-voltage vircator (Fig. 1) contains an electron gun *1* generating an axisymmetric converging electron beam *2*, which was injected into a diode system comprising two grid electrodes forming a retarding field. The retarding field was created by applying a negative potential $V_r$ to the exit (second) grid *4* relative to the entrance (first) grid *3*. The accelerating voltage in our experiments was 3.0 kV, the electron beam current was $I_0 = 250$ mA, the distance between grids of the diode gap was $l = 50$ mm, and the beam radius was $r_b = 4$ mm. Past the interaction region, the electron beam strikes collector 5. The experiments were performed in a continuously evacuated vacuum setup at a minimum residual pressure of $10^{-7}$ Torr.

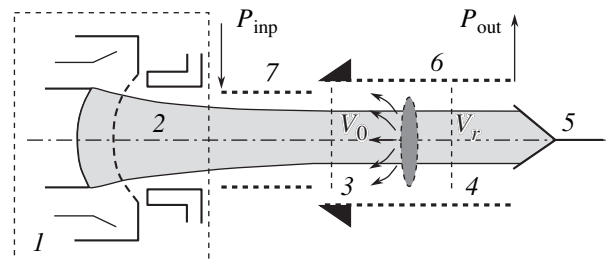

**Fig. 1.** Schematic diagram of the experimental setup used to study the effect of external signal on the generation of oscillations in a nonrelativistic electron beam with virtual cathode: (*1*) electron gun; (*2, 3*) input grid system; (*4*) output grid to which the retarding potential $V_r$ is applied; (*5*) electron collector; (*6*) helical slow-wave system segment (microwave power extraction); (*7*) helical slow-wave system segment for the introduction of an external signal.

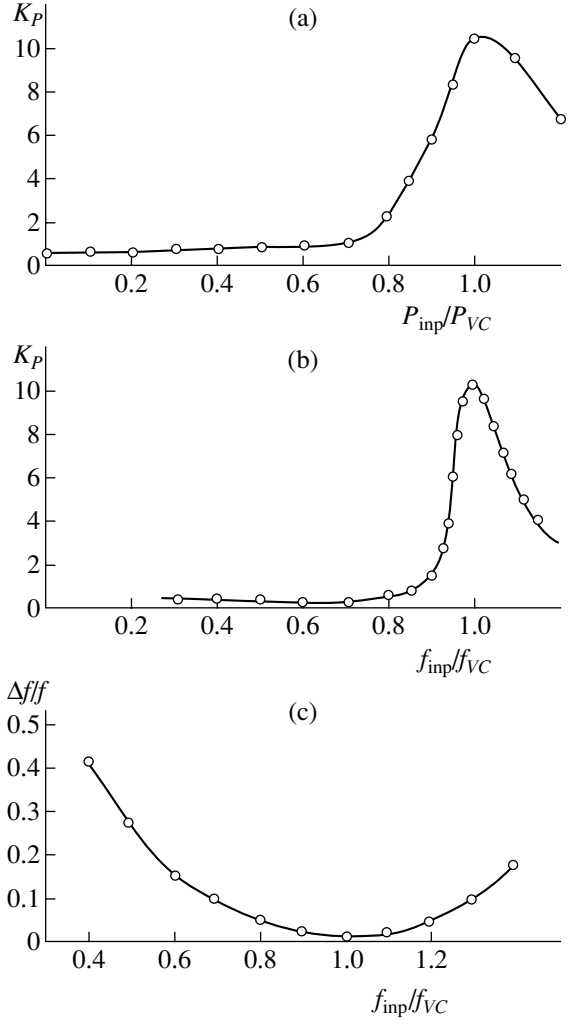

**Fig. 2.** Plots of the (a) power gain $K_P = P_{out}/P_{inp}$ in the beam with VC versus normalized external signal power $P_{inp}/P_{VC}$ at $f_{inp} \approx f_{VC}$; (b) power gain $K_P$ versus normalized external signal frequency $f_{inp}/f_{VC}$ at $P_{inp} \approx P_{VC}$; (c) output signal bandwidth $\Delta f/f$ versus normalized external signal frequency $f_{inp}/f_{VC}$ at $P_{inp} \approx P_{VC}$.

As the retarding potential difference $\Delta V_r = |V_0 - V_r|$ between grids of the diode gap is increased to a certain level, the electron beam is retarded and an oscillating VC is formed in the system. The radiation generated by this oscillator is extracted via a broadband helical segment 6. A further increase in the retarding potential difference $\Delta V_r$ is accompanied by a transition from regular to chaotic oscillations of the VC [6]. In this study, the retarding potential wad maintained on a constant level such that $\Delta V_r/V_0 = 0.35$.

The external signal was introduced via an additional slow-wave system segment 7 that was situated between the second anode, electron gun, and the first grid (see Fig. 1). Application of the external signal leads to modulation of the elevation beam at the entrance of the interaction space of the low-voltage vircator. The external signal was generated by G4-79 and G4-80 oscillators. In autonomous regimes (in the absence of the external signal) the low-voltage vircator generated a narrowband microwave signal with a base frequency of $f_{VC} \approx 2.0$ GHz, a bandwidth of $\Delta f/f \sim 0.5\%$, and an output power of $P_{VC} = 20$ mW.

We have studied the influence of the power ($P_{inp}$) and frequency ($f_{inp}$) of the external signal on the output characteristics of the low-voltage vircator, in particular, on the output signal power $P_{out}$ and bandwidth of $\Delta f/f$. Figure 2 shows a plot of the ratio $K_P = P_{out}/P_{inp}$ of the output and input signal powers (vircator power gain) versus external signal power $P_{inp}/P_{VC}$ normalized to the vircator power in the autonomous generation regime. The external signal frequency was equal to that of the free VC oscillations: $f_{inp} = f_{VC}$. As can be seen from Fig. 2a, there is a small suppression of the output signal at low levels of the external signal power ($K_P \approx 0.4$). As the external action is increased, $K_P$ exhibits growth and, at an external signal power close to that of the autonomous generation ($P_{inp} \approx P_{VC}$), the gain reaches a level of $K_P > 10$ dB, which corresponds to a vircator output power of about 200 mW. Thus, we may speak of a relatively large signal amplification in the electron-wave system with VC for an external signal frequency close to that of the VC oscillations. Note that the observed effect cannot be explained in terms of the classical synchronization of autooscillations in the beam with VC [15], since a more than tenfold power gain is observed in the system.

Let us consider the power gain as dependent on the frequency of the external signal. Figure 2b shows a plot of $K_P$ versus external signal frequency $f_{inp}/f_{VC}$ normalized to the VC frequency in the autonomous generation regime. At large detunings $|f_{inp} - f_{VC}| << 0.9$, the external signal is also suppressed, but, on approaching the VC frequency ($f_{inp} \sim f_{VC}$), the power gain exhibits a sharp increase. Therefore, the low-voltage vircator operates as a narrowband active filter with a gain on the order of 10 dB and bandwidth of $\Delta f/f_{VC} \approx 0.1$ (on a 3 dB level).

It should also be noted that variation of the external signal frequency leads to a change in the spectrum of microwave generation of the low-voltage vircator. Fig. 2c shows a plot of the output signal bandwidth $\Delta f/f$ versus normalized external signal frequency at $P_{inp} \approx P_{VC} f_{inp}/f_{VC}$. As can be seen, large detunings of the external signal frequency from that of VC in the autonomous generation regime lead to the generation of a broadband chaotic signal with $\Delta f/f \approx 0.2$–$0.4$. As the relative detuning decreases, the generation bandwidth exhibits monotonic narrowing and the case of $\Delta f/f \approx 1$ corresponds to VC oscillations in a nearly single-frequency regime. Thus, the power amplification of nonautonomous oscillations is accompanied by the transition from a chaotic broadband oscillations to a narrowband generation regime close to the regular VC oscillations.

In conclusion, we have experimentally demonstrated the possibility of a significant (by more than 10 dB) increase in the power output of a low-voltage vircator under the action of an external signal. The power gain is observed when the external signal frequency is close to the VC frequency. The effect exhibits a threshold character with respect to the external signal amplitude. The maximum output signal power is reached at an external signal power close to that of the autonomous generation.

The observed effect can be used for the amplification of microwave signals in systems featuring charged particle beams with VCs. Advantages of the proposed amplifier are the simple design, the possibility of operation without a guiding magnetic field, and the possibility of tuning the frequency of amplified signals in a broad band by varying the retarding potential applied to the second grid (electron beam perveance in the intergrid gap) [6]. It should be noted that the frequency of autonomous VC oscillations is proportional to the plasma frequency of the electron beam ($f_{KC} \sim f_p$) (see, e.g., [1, 5, 6]). This implies that an increase in $f_p$ of the beam due to a small increase in the electron density (e.g., with the aid of a magnetron injector gun [7, 16]) makes it possible to move toward the millimeter wavelength range with prospects of creating a simple and compact low-voltage microwave amplifier based on the electron beam with virtual cathode.

**Acknowledgments.** The authors are grateful to Prof. D.I. Trubetskov for his interest in this study, fruitful discussions, and useful critical remarks.

This study was supported by the Russian Foundation for Basic Research (project nos. 05-02-16273, 05-02-16286). A.E.H. gratefully acknowledges support from the "Dynasty" Noncommercial Program Foundation and the US Civilian Research and Development Foundation for the Independent States of the Former Soviet Union (CRDF award no. Y2-P-06-06).